%% file: main.tex
\documentclass[aps,prc,twocolumn,superscriptaddress,nofootinbib,preprintnumbers]{revtex4-1}

\usepackage[utf8]{inputenc}

\usepackage{mathtools}
\usepackage{amssymb}
\usepackage{amsfonts}
\usepackage{mathrsfs}
\usepackage{bbm}
\usepackage{slashed}
\usepackage{xspace}

\usepackage[colorlinks=true,linkcolor=blue,citecolor=blue, urlcolor=blue]{hyperref}
\usepackage[nameinlink]{cleveref}
\Crefname{equation}{Eq.}{Eqs.}
\Crefname{figure}{Fig.}{Figs.}
\Crefname{tabular}{Tab.}{Tabs.}
\Crefname{section}{Sec.}{Secs.}
\Crefname{appendix}{Appendix}{Appendices}
\usepackage{bookmark}

\usepackage{xcolor}
\usepackage{enumitem}
\hypersetup{colorlinks   = true, %Colours links instead of ugly boxes
	urlcolor     = blue, %Colour for external hyperlinks
	linkcolor    = blue, %Colour of internal links
	citecolor   = blue}

\usepackage{units}

\usepackage[utf8]{inputenc}

\usepackage{soul}

\newcommand{\dd}{\mathrm{d}}

\newcommand{\zz}{\mathbb{Z}}
\newcommand{\rr}{\mathbb{R}}

\newcommand{\Pcal}{\mathcal{P}}

\newcommand{\SU}{\mathrm{SU}}

\newcommand{\trajectum}{\emph{Trajectum}} % DS: I've deleted ~ after Trajectum, since it results in weird spacings if Trajectum is written at sentence endings

\newcommand{\fivenn}{$\sqrt{s_{\mathrm{NN}}}=5.02$~TeV\xspace}
\newcommand{\PbPb}{\mbox{Pb--Pb}\xspace}
\newcommand{\OO}{\mbox{O--O}\xspace}

\input{commands.tex}

\newcommand{\getMPIAffiliation}{\affiliation{Max Planck Institute for Mathematics in the Sciences, Inselstraße 22, Leipzig, 04103, Germany}}
\newcommand{\getHeidelbergPIAffiliation}{\affiliation{Physical Institute, Heidelberg University, Im Neuenheimer Feld 226, 69120 Heidelberg, Germany}}
\newcommand{\getGSIAffiliation}{\affiliation{GSI Helmholtz Centre for Heavy Ion Research GmbH, Planckstraße 1, 64291 Darmstadt, Germany}}
\newcommand{\getCERNAffiliation}{\affiliation{Theoretical Physics Department, CERN, CH-1211 Gen\`eve 23, Switzerland}}

\begin{document}

\title{Towards a topological data analysis for heavy-ion collisions}

\author{Federica Capellino}
\email{f.capellino@gsi.de}
\getGSIAffiliation

\author{Andrea Dubla}
\email{a.dubla@gsi.de}
\getGSIAffiliation

\author{Silvia Masciocchi}
\email{s.masciocchi@gsi.de}
\getHeidelbergPIAffiliation
\getGSIAffiliation

\author{Govert Nijs}
\email{govert.nijs@cern.ch}
\getCERNAffiliation

\author{Daniel Spitz}
\email{daniel.spitz@mis.mpg.de}
\getMPIAffiliation

\begin{abstract}
The collective expansion of the quark-gluon plasma (QGP) created in heavy-ion collisions suggests that geometry-inspired approaches can be useful in extracting information about the QGP. 
In this work, a systematic study of observables based on topological data analysis is provided for simulations of heavy-ion collisions.
Specifically, we implement persistent homology observables for metric-based complexes in the heavy-ion model \emph{Trajectum} and provide predictions for \PbPb and \OO collisions, where the tunable model parameters are taken from a Bayesian analysis performed in \PbPb collisions \cite{Giacalone:2023cet}\@.
This, in particular, allows us to compute systematic uncertainties on our observables from the uncertainties in the model parameters.
To bridge between new and already established observables, we build a dictionary linking the topological observables to traditional ones, such as particle multiplicities, momentum distributions, and the elliptic flow coefficient. 
While the persistent homology observables largely reflect known phenomenology and do not show enhanced sensitivity to the model's tunable parameters compared to conventional observables, this study demonstrates the viability and robustness of topological techniques in the context of heavy-ion physics. 
They may offer alternative perspectives and potential applications in heavy-ion physics.
\end{abstract}

\preprint{CERN-TH-2025-168}
\maketitle

%%%%%%%%%%%%%%%%%%%%%%%%%%%%%%%%%%%%%%%%%%%%%%%%%%%%%%%%%%%%%%%%%%%%%%%%
\section{Introduction}
%%%%%%%%%%%%%%%%%%%%%%%%%%%%%%%%%%%%%%%%%%%%%%%%%%%%%%%%%%%%%%%%%%%%%%%%
High-energy heavy-ion collisions at the Relativistic Heavy Ion Collider (RHIC) and the Large Hadron Collider (LHC) produce a deconfined state of quarks and gluons, called quark-gluon
plasma (QGP) 
~\cite{Busza:2018rrf, ALICE:2022wpn, STAR:2005gfr, PHENIX:2004vcz}. 
In the past decades, hydrodynamic models, including viscosity, have been applied with great success to describe the distribution of soft hadrons produced in heavy-ion collisions. 
While first principles calculations of the macroscopic fluid properties are challenging, phenomenological and theoretical studies are motivated by an increasing amount of experimental results.
In recent years, the large wealth of experimental
data has been used to provide Bayesian estimates for
the transport coefficients. Multi-observable model-to-data fits focused on the transverse momentum (\pt) and centrality dependence of several observables, such as particle multiplicity or anisotropic flow coefficients~\cite{Nijs:2020roc,Nijs:2020ors,Schenke:2020mbo,Vermunt:2023zsk,Lu:2024upk,Heffernan:2023utr,Heffernan:2023gye,JETSCAPE:2020shq,Parkkila:2021yha,Virta:2024avu}\@. Furthermore, in recent years great attention was devoted to smaller systems such as oxygen-oxygen (\OO) and neon-neon (Ne--Ne) collisions to study the applicability of hydrodynamics in these small collision systems \cite{Giacalone:2024ixe, Giacalone:2024luz,Brewer:2021kiv}\@. 
With the ongoing run 3 at the LHC, the available statistics for lead–lead (\PbPb) collisions will increase significantly, and new data were collected from \OO and Ne–Ne collisions. These developments provide strong motivation to explore novel observables.

In this work, we perform a systematic study of topological data analysis (TDA), providing new geometry-based observables for heavy-ion collisions.
TDA can provide computational methods to sensitively study the intricate connectivity structures emerging from the final state particle momenta and might as such form valuable new observables with parametrically discriminative power.
Persistent homology, the prevailing TDA tool, allows for the extraction of scale-dependent, robust topological features such as connected components and holes appearing in a sequence of spaces (complexes) inferred from the event data~\cite{otter2017roadmap, ChazalMichelIntro}\@.
In the realm of heavy-ion collisions, it has been shown that persistent homology can provide comprehensive clustering signatures related to long-range flow correlations~\cite{Hamilton:2022blu}\@.

Topological machine learning has been applied to analyze critical fluctuations in intermittency 
during heavy-ion collisions, aiming at distinguishing weak signal events from background noise in persistent homology~\cite{Wang:2024bzy}\@.
For highly energetic jets, persistent homology allowed for topology-leveraging jet tagging~\cite{thomas2022topological}\@.
The underlying complexes have been explored as informative representations of jet substructure~\cite{Gaertner:2023ycs}\@.
In the related context of lattice gauge theories, persistent homology has been applied to study universal dynamics in a gluonic plasma~\cite{Spitz:2023wmn} as well as the confining and deconfining phases of pure $\SU(2)$ and $\SU(3)$ gauge theories~\cite{Spitz:2022tul, Spitz:2024bqh}\@.
Furthermore, persistent homology has been employed to probe strings, center vortices, and monopoles~\cite{Sehayek:2022lxf, Sale:2022qfn, Crean:2024nro}\@.

Data-wise, our approach is based on 2-dimensional point clouds formed by the transverse momenta of charged hadrons in the final states simulated using the \trajectum~relativistic hydrodynamics code~\cite{Nijs:2020roc}\@. 
Persistent homology is then computed for so-called alpha complexes~\cite{edelsbrunner1983shape,edelsbrunner1994three, otter2017roadmap} for the Euclidean 2-norm in the transverse momentum space.
This corresponds to growing balls around the final state particle momenta and studying properties such as the number of connected components or the number of holes of their union.

In the present work we focus on the Betti curves and the distributions of persistence values, for unidentified charged hadrons as well as for $\pi^\pm$, $K^\pm$, and $p$ in five centrality classes (ranging over 0--40\%) of \PbPb collisions at the center-of-mass energy per nucleon pair \sqrtsNN = 5.02~TeV and in \OO collisions at \sqrtsNN = 7\,TeV at the LHC\footnote{Data for \OO collisions have recently been collected at \sqrtsNN = 5.36\,TeV at the LHC\@. Our results were produced prior to that. If persistent homology observables will be measured, the theoretical results for the correct energy will be provided.}\@.
For each persistent homology observable we provide systematic and statistical uncertainties.
The results for \PbPb and \OO collisions are comprehensively compared, providing signals of the vastly different system sizes.
Through this method we reveal furthermore that persistent homology can simultaneously encode multiple
phenomenological features present in traditional heavy-ion collision observables such as particle multiplicities and radial as well as elliptic flows.
Our methodological approach is complementary to the one employed in~\cite{Hamilton:2022blu}, which utilizes more complicated constructions of the complexes.

This paper is structured as follows.
We describe the concepts behind persistent homology in \Cref{persistent homology}\@. 
We summarize the details of the initial conditions, the hydrodynamic evolution, and the hadronization procedures together with the Bayesian inference analysis procedure and determination of its uncertainties in \Cref{trajectum}. 
We then provide the predictions for the persistent homology observables in \Cref{results}\@. 
Finally, we provide a summary and possible future directions in \Cref{conclusion}\@.

\begin{figure*}[t]
    \centering
	\includegraphics[scale = 0.78]{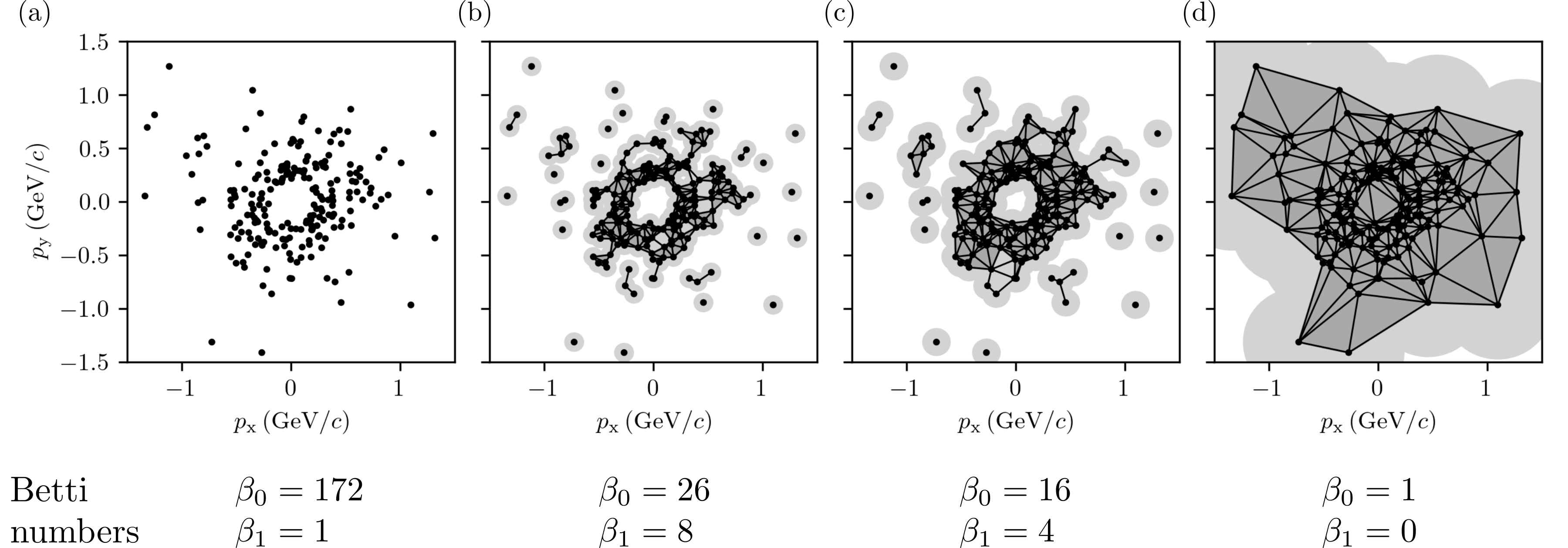}
	\caption{Alpha complexes of a point cloud given by transverse momentum space (\px, \py) positions of produced $\pi^+$
    in a Pb--Pb collision at \sqrtsNN = 5.02 TeV generated by \trajectum. 
 The empty disk in the point cloud center is due to a kinematic selection, particles with \pt $<$ 0.2 GeV$/c$ are not included in the analysis.
 The radii of the alpha complexes are illustrated as grey disks around the data points: (a) $r$=0.02 GeV$/c$, (b) $r$=0.08 GeV$/c$, (c) $r$=0.12 GeV$/c$ and (d) $r$ = 0.5 GeV$/c$\@. 
 The corresponding alpha complexes are shown as black edges and dark grey triangles along with the vertex points. 
 Underneath the alpha complexes, we show the corresponding zeroth and first Betti numbers.
 }
 \label{Fig:AlphaComplexesIllustration}
\end{figure*}

%%%%%%%%%%%%%%%%%%%%%%%%%%%%%%%%%%%%%%%%%%%%%%%%%%%%%%%%%%%%%%%%%%%%%%%%
\section{Concepts of persistent homology}
\label{persistent homology}
%%%%%%%%%%%%%%%%%%%%%%%%%%%%%%%%%%%%%%%%%%%%%%%%%%%%%%%%%%%%%%%%%%%%%%%%

This section is devoted to the introduction of the topological observables used in this work, which allow the study of
connectivity structures in simulations of heavy-ion collisions.
First, we introduce certain combinatorial objects inferred from the simulation data: the family of alpha complexes.
Their topology can be relatively quickly computed using the concept of homology.
Sweeping through the family of alpha complexes, changes in their shape or connectivity (topology) can be monitored using persistent homology, which is subsequently described.

For better readability, we focus on intuitive descriptions of the employed mathematical objects here, restricting to a minimum of mathematics.
\Cref{Appendix:Homology} provides a brief description of the underlying mathematical constructions, which lead to homology and persistent homology.
We refer to the literature for more comprehensive introductions, see e.g.~\cite{otter2017roadmap, ChazalMichelIntro}\@.

\subsection{Alpha complexes}
Simulations of high-energy heavy-ion collisions (see \Cref{trajectum} for details) are carried out to generate point clouds of produced particles in the 2-dimensional transverse momentum space with coordinates \pt$=(p_x, p_y)$\@.
Each point corresponds to a final state particle of the event.
The topology of the point clouds themselves is rather trivial and merely amounts to point counting.
Instead, more advanced notions of length scale-dependent topology can be inferred from such point cloud data.
For this, one first constructs combinatorial objects from them, so-called simplicial complexes, whose topology one then studies.
A simplicial complex is a collection of triangles of different dimensions (called simplices), closed under taking boundaries.
For instance, a dimension-0 simplex is a point, a dimension-1 simplex is a line between two points, a dimension-2 simplex is a triangle.
The boundaries of a dimension-2 simplex are its three bounding edges, the boundary of a dimension-1 simplex is given by its two endpoints and the boundary of a dimension-0 simplex is defined to be empty.

We focus on alpha complexes, which are among the computationally most efficient and informative simplicial complexes~\cite{edelsbrunner1983shape,edelsbrunner1994three, otter2017roadmap}\@.
With regard to their topology, their construction can be described as follows.\footnote{The described construction actually yields the 2-skeleton of \v{C}ech complexes.
For alpha complexes, the construction is more elaborate and uses circumcircles of minimal radii inscribed in the point cloud to determine the triangles.
Yet, under very general conditions \v{C}ech and alpha complexes have the same persistent homology~\cite{bauer2017morse}, so that in this regard the two can be used interchangeably.} 
We let balls of radius $r$ grow around the particle positions \pt in momentum space, from which the complexes are defined.
First, for all radii $r\geq 0$ the complexes comprise the point cloud itself.
When two of the balls intersect for a particular radius $r$, an edge between the corresponding center points is included in the complex.
When, for some $r$, three balls have a non-empty intersection, a triangle is included in the complex, inscribed between the three center points.
In the two dimensions of the transverse momentum space, these simplices are all that can occur.

Alpha complexes of different radii are illustrated in the top row of \Cref{Fig:AlphaComplexesIllustration} for an example event generated using \trajectum\@.
The indicated positions are transverse momentum space positions of produced final state $\pi^+$.
The empty disk in the center of the point cloud is due to a kinematic selection of \pt$ \geq 0.2~\text{GeV}/c$, which is employed before running the analyses to be consistent with the momentum region accessible at the experiments.
\Cref{Fig:AlphaComplexesIllustration}(a) shows the alpha complex for a small radius $r = 0.02~\text{GeV}/c$ at which the balls barely intersect and the alpha complex is visually dominated by the individual points.
For a larger radius, edges between the points appear, see \Cref{Fig:AlphaComplexesIllustration}(b) for $r = 0.08~\text{GeV}/c$\@.
Upon further increasing $r$, more and more points become connected by edges and many triangles appear, see \Cref{Fig:AlphaComplexesIllustration}(c) for $r = 0.12~\text{GeV}/c$\@.
Finally, for sufficiently large radii, the alpha complex becomes independent of $r$ and comprises all triangles, see \Cref{Fig:AlphaComplexesIllustration}(d)\@.

\subsection{Persistent homology}\label{SecPersHom}
The topology of the alpha complexes can be suitably described employing homology, which provides information on the connected components as well as the holes, and is invariant under continuous deformations of the complexes.
In homology, connected components and holes are described by means of homology classes, which can have different homology dimensions, see also \Cref{Appendix:Homology}.
For instance, a dimension-0 homology class corresponds to an individual connected component, and a dimension-1 homology class provides loop-like paths through the alpha complex, which are not fully filled with triangles, i.e., it describes a hole in the complex.

In the two dimensions of the transverse momentum space, these are all types of non-trivial homology classes, which can occur.
Moving through the family of alpha complexes as in \Cref{Fig:AlphaComplexesIllustration}(a) through (d), we see that a multitude of dimension-0 and dimension-1 homology classes can appear for a generic event.

Crucially, \Cref{Fig:AlphaComplexesIllustration} suggests that the topology of the alpha complexes changes with the radius $r$\@.
For instance, for low radii as in \Cref{Fig:AlphaComplexesIllustration}(a) many independent connected components appear (dimension-0 homology classes), since many of the points have not yet been connected by edges at such small radii.
This only happens at somewhat larger radii as in \Cref{Fig:AlphaComplexesIllustration}(b), where quite a few of the connected components have merged (i.e., they \emph{died})\@.
In particular, for this radius also the first holes have already formed in the alpha complex (they \emph{got born}), so that non-trivial dimension-1 homology classes appear and \emph{persist} in the family of alpha complexes for a certain radius interval.
This trend continues towards larger radii, see e.g.~\Cref{Fig:AlphaComplexesIllustration}(c).
Finally, for sufficiently large radii, all holes have been filled entirely with triangles, i.e., they all died, and we are left with a single connected component without holes, see \Cref{Fig:AlphaComplexesIllustration}(d)\@.

For a given alpha complex, the independent \mbox{dimension-$\ell$} homology classes are counted by the $\ell$-th Betti number $\beta_\ell$ for $\ell=0,1$\@.
For the alpha complexes shown in \Cref{Fig:AlphaComplexesIllustration}, the Betti numbers $\beta_0$ and $\beta_1$ are given underneath the complexes.
They clearly follow the qualitative behavior outlined in the previous paragraph.
The \emph{Betti curves} $\beta_\ell(r)$ are defined as the $\ell$-th Betti numbers of the alpha complex at radius $r$ and describe their radius dependence.
While many persistent homology descriptors other than the Betti curves are available~\cite{hensel2021survey}, due to their geometric interpretability and numerical accessibility we mainly focus on these in the present paper.

In addition, we consider the distribution of persistence values $\dd N/\dd \mathcal{P}_\ell$ of dimension-$\ell$ homology classes.
The persistence of a dimension-$\ell$ homology class is defined as $\mathcal{P}_\ell:=r_d-r_b$, where $r_d$ is the death radius of the homology class and $r_b$ is its birth radius\footnote{Other than the linear persistence $\Pcal_\ell = r_d-r_b$, the death-birth ratio $r_d/r_b$ also provides a measure for the dominance of homological features in the alpha complexes. Crucially, the ratio is by construction scale-insensitive. Yet, it can be viewed as less informative, since it has been shown that its distribution is universal across many point cloud generation processes~\cite{bobrowski2022universality, bobrowski2024weak}\@.}.
Intuitively, persistences provide measures of the visual dominance of a topological feature.
For instance, the connected components corresponding to the outlier points in \Cref{Fig:AlphaComplexesIllustration} are born at radius $r_b=0$ but only die at radius half the distance to their nearest neighbors, resulting in comparably large persistence values.
An example for a dimension-1 homology class with large persistence is provided by the empty disk in the center of the point cloud of \Cref{Fig:AlphaComplexesIllustration}.
Already in \Cref{Fig:AlphaComplexesIllustration}(b) the corresponding homology class is present and persists up to a radius between the ones of \Cref{Fig:AlphaComplexesIllustration}(c) and \Cref{Fig:AlphaComplexesIllustration}(d).

Persistent homology has advantageous properties. It can be straight-forwardly computed using modern algorithms, evaluating the homology and related birth-death radii for the whole family of alpha complexes at once.
We employ the computational topology library GUDHI in C++~\cite{boissonnat2014gudhi}\@.
Furthermore, facilitating numerical approaches such as ours, persistent homology is provably stable against perturbations of the input point cloud, see e.g.~\cite{cohen2007stability, cohen2010lipschitz}.
Finally, persistent homology and the Betti curves can be well analyzed in statistical contexts~\cite{hiraoka2018limit, spitz2020self}\@.

%%%%%%%%%%%%%%%%%%%%%%%%%%%%%%%%%%%%%%%%%%%%%%%%%%%%%%%%%%%%%%%%%%%%%%%%
\section{Modeling of heavy-ion collisions and Bayesian inference analysis}
\label{trajectum}
%%%%%%%%%%%%%%%%%%%%%%%%%%%%%%%%%%%%%%%%%%%%%%%%%%%%%%%%%%%%%%%%%%%%%%%%

\begin{figure*}[ht!]
    \centering
    \includegraphics[width=0.48\linewidth]{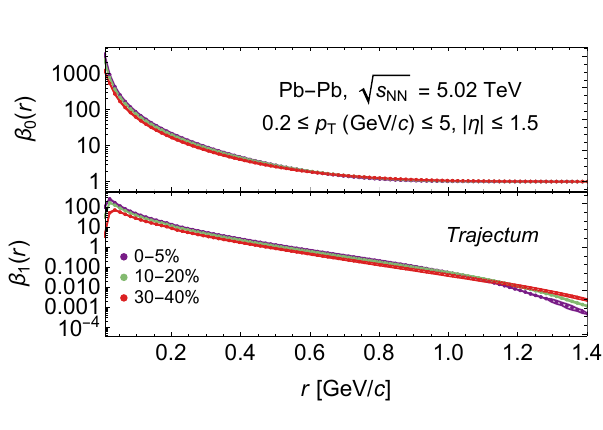}
    \includegraphics[width=0.48\linewidth]{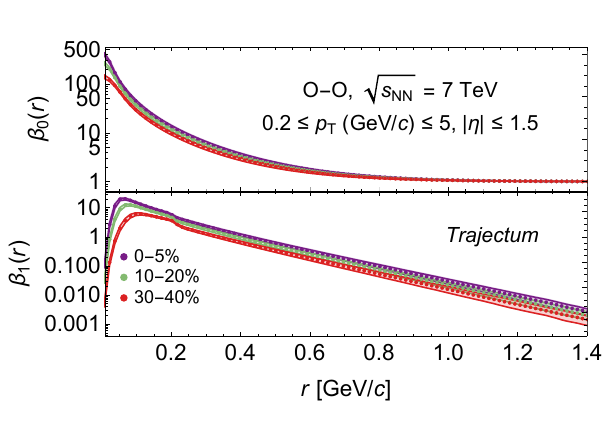}
    \caption{Dimension-0 and 1 Betti curves for \PbPb (left) and \OO collisions (right) at \fivenn and \sevennn, respectively.
    Three different centralities from head-on to mid-central collisions are shown.
    The curves reveal a twofold multiplicity dependence: more peripheral collisions result in fewer homological features, which are furthermore larger in momentum space extent due to the sparser event point clouds.   }
    \label{fig:betti_centrality}
\end{figure*}
We use the \trajectum~framework as in~\cite{Nijs:2020roc, Nijs:2020ors}\@.
In \emph{Trajectum}, we model a collision by first generating an initial state based on the nucleon positions, using a model that generalizes the T\raisebox{-0.5ex}{R}ENTo model~\cite{Moreland:2014oya}\@.
This is then followed by a far-from-equilibrium stage that interpolates between free streaming and a holography-inspired scenario~\cite{Nijs:2023yab}\@.
The result of this then feeds into a simulation of the hydrodynamic phase, and as the fluid cools, at a fixed freeze-out temperature, we switch to a particle-like description.
The resulting particles are evolved using the SMASH code~\cite{SMASH:2016zqf, dmytro_oliinychenko_2020_3742965}\@.

The model described above has many parameters, the values of which are hard to determine theoretically.
To obtain reasonable values for these parameters, in~\cite{Giacalone:2023cet} a fit to data was performed, where 26 model parameters were varied (for further details see \Cref{Appendix:parameters})\@.
The observables used in that fit include particle yields, mean transverse momenta, anisotropic flow and fluctuations of mean transverse momentum, but also \pt-differential observables such as spectra and \pt-differential flow.
Finally, several `statistically difficult' observables such as the $\rho(v_2\{2\}^2,\langle p_\mathrm{T} \rangle)$ and $NSC(2, 3)$ and $NSC(2, 4)$ observables were included in the fit.
Crucially for this work, no persistent homology observables were included.
Each observable used in the fit is computed using the exact same kinematic selections as the corresponding experimental measurement, making for a comparison that is as apples-to-apples as possible.
For full details, see~\cite{Giacalone:2023cet}\@.

In this work, we use the parameters obtained in the fit described above.
However, we do not just take the most likely parameters.
Instead, we take 20 parameter choices, which are randomly sampled from the posterior.
As such, these choices represent the residual uncertainty in these parameters after fitting to the data.
By performing around 400k calculations for each of these different choices, we can not only obtain a central value and a statistical uncertainty for our prediction, but also a systematic uncertainty, which can be seen as the propagated uncertainty in the parameters.

We compute our observables for each of these 20 parameter choices, thereby generating an ensemble of predictions. The spread in this ensemble, encoded in the standard deviation $\sigma_\text{tot}$, represents the total uncertainty of the observable. However, some of this total is due to the statistical uncertainty of each of the 20 model computations, denoted by $\sigma_\text{stat}$\@. Since variances are additive, we compute the systematic uncertainty as
\[
\sigma_\text{syst} = \sqrt{\sigma_\text{tot}^2 - \sigma_\text{stat}^2}.
\]

The fit in \cite{Giacalone:2023cet} fits only to \PbPb observables.
To perform calculations for \OO collisions, we use explicit configurations computed using nuclear lattice effective field theory (NLEFT) \cite{Giacalone:2024luz}\@.

%%%%%%%%%%%%%%%%%%%%%%%%%%%%%%%%%%%%%%%%%%%%%%%%%%%%%%%%%%%%%%%%%%%%%%%%
\section{Results}
\label{results}

%%%%%%%%%%%%%%%%%%%%%%%%%%%%%%%%%%%%%%%%%%%%%%%%%%%%%%%%%%%%%%%%%%%%%%%

In this section, the results for Betti curves and persistence distributions are discussed.
The study consists of results for \PbPb and \OO collisions at center-of-mass collision energies \fivenn and \sevennn, respectively. Each displayed Betti curve is the result of an average over all events in a centrality class.
First, we focus on the dependence of the Betti curves on centrality and the momentum radius parameter $r$, followed by a study of Betti curves for different particle species. In particular, we highlight the nontrivial influence of angular correlations on these observables and we point to the analogies with traditional heavy-ion observables.
We then investigate the persistence distributions, which we find to be sensitive to the specific geometry of final-state particle point clouds.
Finally, we systematically investigate which model parameters the Betti curves are sensitive to.

\subsection{Centrality dependence}

In \Cref{fig:betti_centrality} the Betti curves for homology dimensions zero and one are shown as a function of momentum space radius for a set of centrality classes, taking all charged hadrons into account.
Results for both \PbPb and \OO are reported in the left and right panels, respectively, for the centrality classes 0--5\% (purple markers), 10--20\% (green markers), and 30--40\% (red markers). Each curve displays error bars, representing the statistical uncertainty, and an error band, representing the systematic uncertainty of the model calculations.

We note that $\beta_0(r)$ at radius $r=0$~\GeVc can be identified with the number of produced particles in the generated events.
This gives rise to the observed centrality ordering of the Betti curves, since the multiplicity of charged hadrons decreases with decreasing centrality.
For \OO the particle multiplicities are on average predicted to be a factor of 10 
smaller than for \PbPb, which can also be seen in the dimension-0 Betti curves.

The dimension-0 Betti curves for \PbPb and \OO decrease monotonically for increasing radii.
This is a result of the construction of persistent homology, since more and more distinct connected components become connected upon increasing the radius, see also the discussion in \Cref{SecPersHom}\@.
Furthermore, the decrease per radius is smaller for \OO than for \PbPb due to a larger point density for the latter, reflecting the larger multiplicity produced with large nuclei, thus yielding on average larger features for \OO collisions.

The dimension-1 Betti curve $\beta_1(r)$ first increases, then decreases for growing radii.
This is as well a geometric effect as outlined in \Cref{SecPersHom}: multiple dimension-0 homology classes first need to merge in order to form a dimension-1 feature.
The resulting peak height again reveals the centrality- and nucleus type-dependent multiplicities: holes are on average more abundant for point clouds comprising more data points.
The kink in $\beta_1(r)$ for radii $r\simeq 0.2$~\GeVc is due to the employed kinematic selection on the \pt of the particles. Particles with \pt $<$ 0.2 \GeVc are not considered in the analysis.
This leaves an imprint in the average death radii through the holes in the point clouds, which also enter the Betti curves.

\begin{figure*}[t]
    \centering
    \includegraphics[width=0.48\linewidth]
    {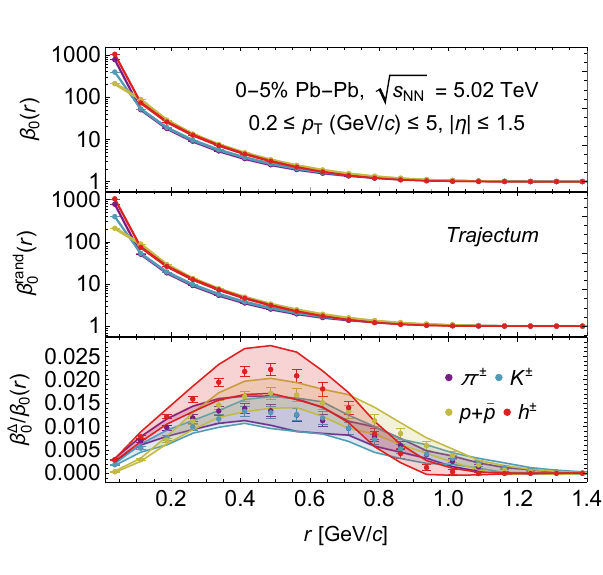}
    \includegraphics[width=0.48\linewidth]{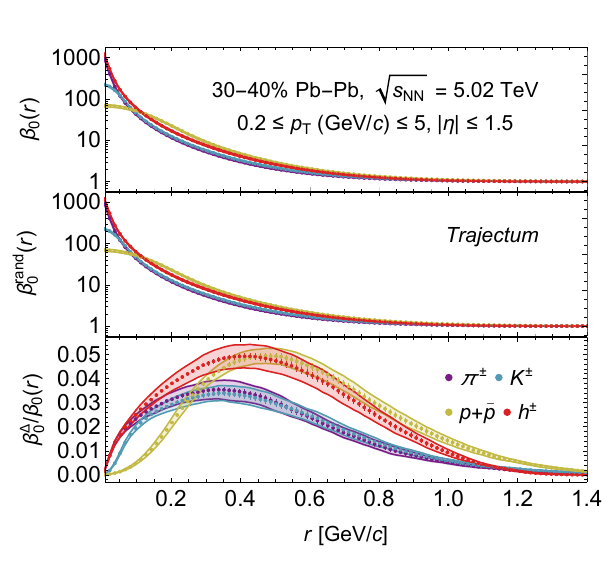}   
    \caption{
    Dimension-0 Betti curves for pions, kaons, (anti)protons and all charged hadrons in the 0--5\% (left panel) and 30--40\% (right panel) centrality class of collisions of \PbPb nuclei.
    The top row provides the plain Betti curves $\beta_0(r)$, the central row the ones computed for events with randomized azimuthal angles, $\beta_0^{\mathrm{rand}}(r)$, and the bottom row provides the ratio $\beta_0^\Delta(r)/\beta_0(r)$, where $\beta_0^\Delta(r) \equiv \beta_0(r)-\beta_0^{\mathrm{rand}}(r)$.
    The curves reflect the mass ordering of the particles and highlight the nontrivial impact of azimuthal correlations on the Betti curves.    }
    \label{fig:betti0_Pb}
\end{figure*}

\subsection{Betti curves}
We now focus on the Betti curves of different identified particle species, specifically charged pions, kaons, and (anti)protons, and examine the influence of azimuthal correlations on the persistent homology results. 
We introduce the Betti curve $\beta_\ell^{\mathrm{rand}}(r)$, which is calculated for the same momentum space point clouds as $\beta_\ell(r)$ but with randomized azimuthal angles. 
That is, for $\beta_\ell^{\mathrm{rand}}(r)$ each particle is assigned an azimuthal angle drawn from a uniform distribution over the interval $[0, 2\pi)$\@.
In addition, we define the quantity
\begin{equation}
    \beta_\ell^\Delta(r) \equiv \beta_\ell(r)-\beta_\ell^{\mathrm{rand}}(r).
\end{equation}
Studying the ratio $\beta_\ell^\Delta(r)/\beta_\ell(r)$ enables us to identify the impact of angular correlations among the particles produced in each event on the Betti curves.

\begin{figure}
    \centering
    \includegraphics[width=\linewidth]{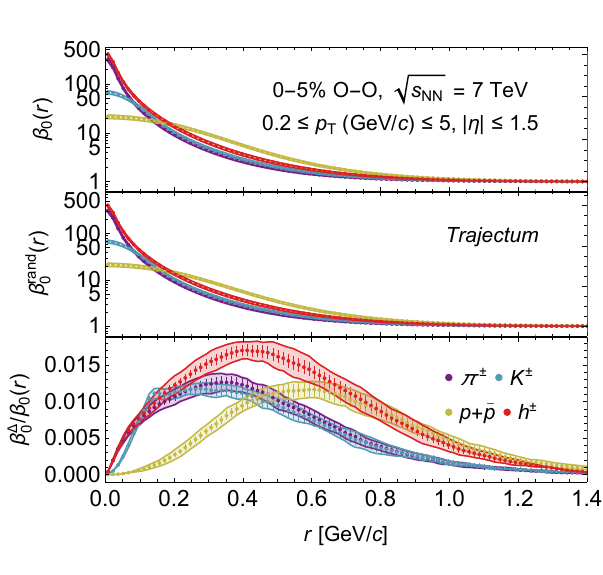}
    \caption{Dimension-0 Betti curves for pions, kaons, (anti)protons and all charged hadrons in the 0--5\% centrality class for collisions of \OO nuclei.
    The top row provides the plain Betti curves $\beta_0(r)$, the central row the ones computed for events with randomized azimuthal angles, $\beta_0^{\mathrm{rand}}(r)$, and the bottom row provides the ratio $\beta_0^\Delta(r)/\beta_0(r)$.
    The curves reflect the mass ordering of the particles and highlight the nontrivial impact of azimuthal correlations on the Betti curves, enhanced with respect to the case of central \PbPb collisions.}
    \label{fig:betti0_O}
\end{figure}

The top panel in \Cref{fig:betti0_Pb} shows the dimension-0 Betti curves as a function of momentum space radius for different particle species and for all charged particles in central (0--5\% centrality is reported in the left panel) and semicentral (30--40\% centrality is reported in the right panel) \PbPb collisions.
As before, the $r=0.0$~\GeVc limit of $\beta_0(r)$ corresponds to the particle multiplicity.
The ordering of the species-specific Betti curves at low radii therefore inversely matches the mass ordering of the particles, since particles with a lower mass are more abundantly produced from the QGP: pions have the largest $\beta_0(r=0)$, followed by kaons and protons, both for central and semicentral collisions.

Along with $\beta_0(r)$, we study $\beta_0^{\mathrm{rand}}(r)$ and $\beta_0^\Delta(r)/\beta_0(r)$, see the middle and bottom rows in \Cref{fig:betti0_Pb}\@.
$\beta_0^{\mathrm{rand}}(r)$ closely resembles $\beta_0(r)$, and all previous considerations also apply in this case. 
However, the variable $\beta_0^\Delta(r)/\beta_0(r)$ sheds further light on the impact of the azimuthal correlations on the dimension-0 Betti curves.
First and foremost, for 
radii of the order of 1-5 times the QCD scale $\Lambda_{\textrm{QCD}} \sim 250\, \textrm{MeV}/c$, $\beta_0^\Delta(r)/\beta_0(r)$ is significantly different from 0, highlighting their nontrivial influence.
For all radii the ratio is positive, since the randomization of the azimuthal angles results on average in larger inter-point distances in the events.

Indeed, due to the azimuthal angle randomization, any slightly ellipsoidal shape of a point cloud turns into a more circular disk-like shape with radius approximately the major axis of the original ellipsoidally shaped point cloud.
The area occupied by the point cloud, or rather by its convex hull, therefore increases on average along with the inter-point distances.
The Betti curves $\beta_0^{\mathrm{rand}}(r)$ are thus slightly shifted towards larger radii compared to $\beta_0(r)$, so that their difference is positive due to their monotonous decline.
In addition, a modification of the shape between the two centrality classes might highlight the larger anisotropic flow in non-central collisions.

Differences in $\beta_0(r)$ between \PbPb and \OO collisions can be observed by comparing the left panel of \Cref{fig:betti0_Pb} with \Cref{fig:betti0_O}, in which the dimension-0 Betti curves are shown for the 0--5\% most central \OO collisions. 
Primarily, for any given radius $\beta_0(r)$ is larger for \PbPb collisions than for \OO.
Furthermore, the decline of the Betti curves for increasing radii is steeper for \PbPb than for \OO.
Such differences arise because \PbPb collisions generate higher energy densities than \OO collisions, resulting in more particles produced within the same centrality class.
As for \Cref{fig:betti_centrality}, the point clouds for \PbPb collisions are denser, leading to smaller homological features. This results in a faster decrease in $\beta_0(r)$ for \PbPb compared to \OO across particle species.

When comparing $\beta_0^\Delta(r)/\beta_0(r)$ for \PbPb collisions, as shown in the bottom panels of \Cref{fig:betti0_Pb}, with the corresponding results for central \OO collisions presented in the lower panel of \Cref{fig:betti0_O}, we observe a similar pattern between \OO and \PbPb collisions. 
Observing the results for $\beta_0^\Delta(r)/\beta_0(r)$ for the different identified particle species, we notice the shift of the curves for protons towards larger homology radii when going from central \PbPb collisions to peripheral \PbPb to central \OO collisions. 
This effect can be associated with the well-known interplay between radial and elliptic flow, which causes a stronger shift for particles with a larger mass towards larger momenta.

\begin{figure*}[th!]
    \centering
    \includegraphics[width=0.48\linewidth]{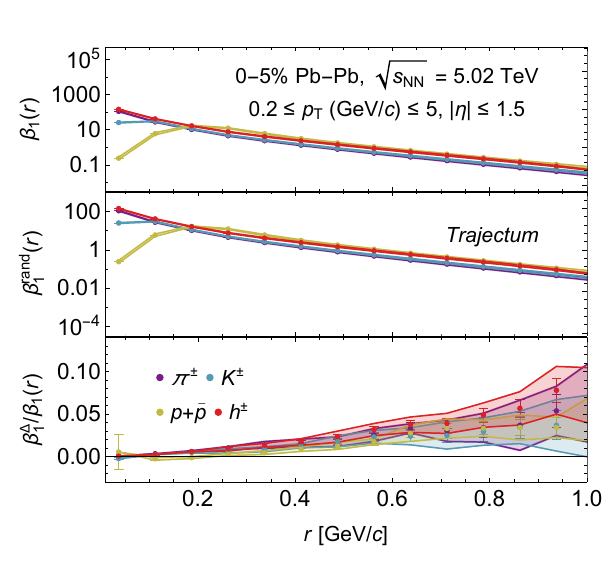}
    \includegraphics[width=0.48\linewidth]{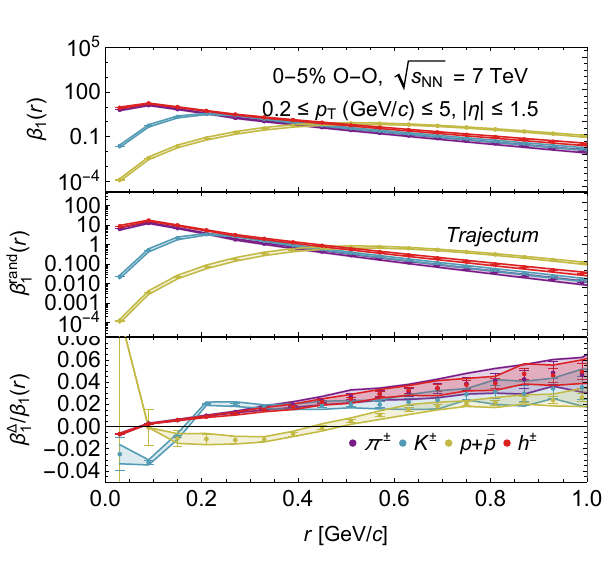}   
    \caption{
    Dimension-1 Betti curves for pions, kaons, (anti)protons and all charged hadrons in the 0--5\% centrality class.
    The left panel shows data for collisions of \PbPb nuclei, the right panel for \OO nuclei.
    The top row provides the plain Betti curves $\beta_1(r)$, the central row the ones computed for events with randomized azimuthal angles, $\beta_1^{\mathrm{rand}}(r)$, and the bottom row provides the ratio $\beta_1^\Delta(r)/\beta_1(r)$\@.
    As in homology dimension 0 (see \Cref{fig:betti0_Pb}), the curves reflect the mass ordering of the particles and highlight the nontrivial impact of azimuthal correlations on the Betti curves.
    }
    \label{fig:betti1_05}
\end{figure*}
In \Cref{fig:betti1_05} the results for the dimension-1 Betti curves are shown as a function of the radius in the 0--5\% centrality class for \PbPb and \OO collisions. 
Due to the lack of statistics at larger homology radii, results for $r>1$ \GeVc are not displayed in the plots. 
The same mass orderings as in the results for homology dimension 0 can be observed: overall Betti numbers are larger for larger particle multiplicities, and so are the homological features in terms of size.
The curves resemble the behavior of the radial flow of different particle species in heavy-ion collisions. 
Analogously to what is observed in \pt distributions of identified charged hadrons, where particles with a larger mass are boosted to higher \pt due to the collective radial expansion of the system, the dimension-1 Betti curves of particles with a higher mass show a peak at a larger momentum space radius.

The impact of azimuthal correlations on the dimension-1 homological features can be studied via $\beta_1^\Delta(r)/\beta_1(r)$\@.
For kaons in \OO collisions the ratio is first negative, then turning positive with a zero crossing around $r\simeq 0.18$~\GeVc\@.
Though less clear, the same effect can be observed for pions, protons and, generally, charged hadrons in \OO collisions.
For \PbPb collisions, the same effect can be observed for very small momentum space radii due to the denser point clouds, although it appears strongly suppressed.
Generally, the zero crossings of the ratio $\beta_1^\Delta(r)/\beta_1(r)$ result from a shift of inter-point distances towards larger values due to the azimuthal angle randomization.
Other than in the already discussed case of homology dimension 0, dimension-1 Betti curves are peaked.
Therefore, a shift of homological features to larger radii approximately results in $\beta^{\mathrm{rand}}_1(r) < \beta_1(r)$ for $r$ below the peak radius and $\beta^{\mathrm{rand}}_1(r)>\beta_1(r)$ for $r$ above the peak radius in the Betti curves.
The locations of the zero crossings roughly match the peak radii in the corresponding Betti curves, see \Cref{fig:betti1_05}.

Interestingly, if we compare $\beta_1^\Delta(r)/\beta_1(r)$ with $\beta_0^\Delta(r)/\beta_0(r)$ as provided in \Cref{fig:betti0_Pb} and \Cref{fig:betti0_O}, we note that overall the impact of azimuthal correlations is larger for dimension-1 features than for dimension-0 features.
This is likely due to the vastly more nontrivial geometric construction of representatives of dimension-1 features, which are furthermore larger on average.

\subsection{Persistence distributions}

\begin{figure*}[ht]
    \centering
    \includegraphics[width=0.48\linewidth]{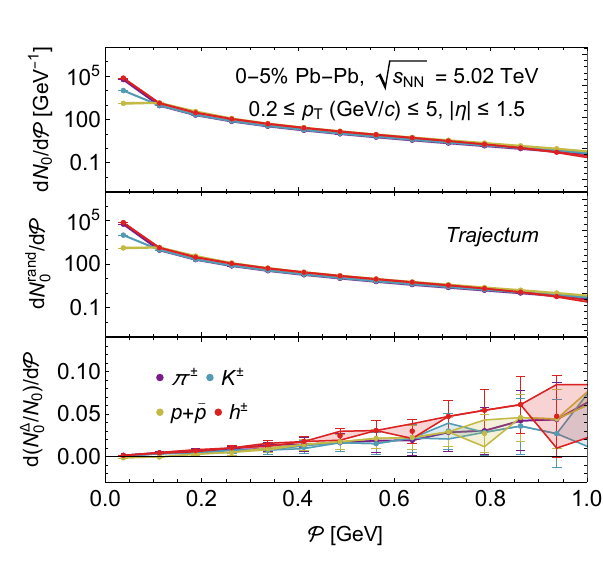}
    \includegraphics[width=0.48\linewidth]{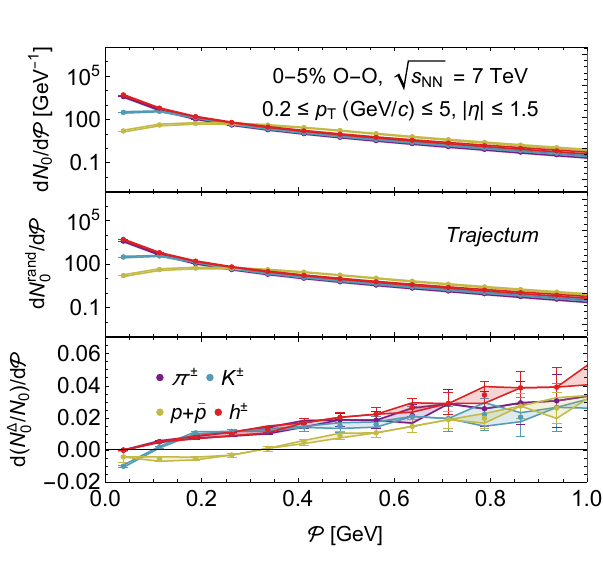}       \caption{Dimension-0 persistence distributions for pions, kaons, (anti)protons and all charged hadrons in the 0--5\% centrality class.
        The left panel shows data for collisions of \PbPb nuclei, the right panel for \OO nuclei.
        The top row depicts the plain persistence distributions $\dd N_0/\dd \Pcal$, the central row has been computed for events with randomized azimuthal angles, $\dd N_0^{\mathrm{rand}}/\dd\Pcal$, and the bottom row provides the ratio $(\dd N_0^\Delta/\dd \Pcal)/(\dd N_0/\dd\Pcal)$.}
    \label{fig:persistencebetti0_05}
\end{figure*}

\begin{figure*}[th!]
    \centering
    \includegraphics[width=0.48\linewidth]{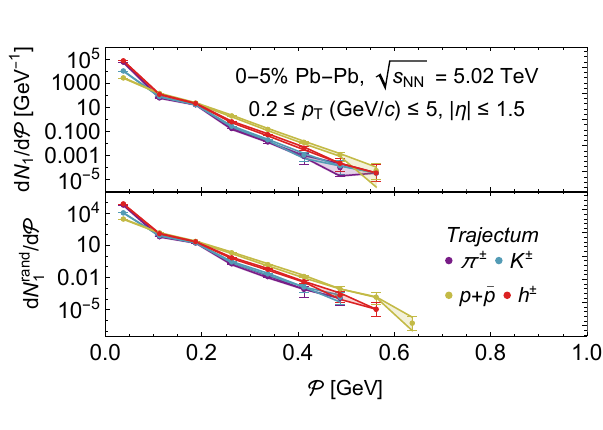}
    \includegraphics[width=0.48\linewidth]{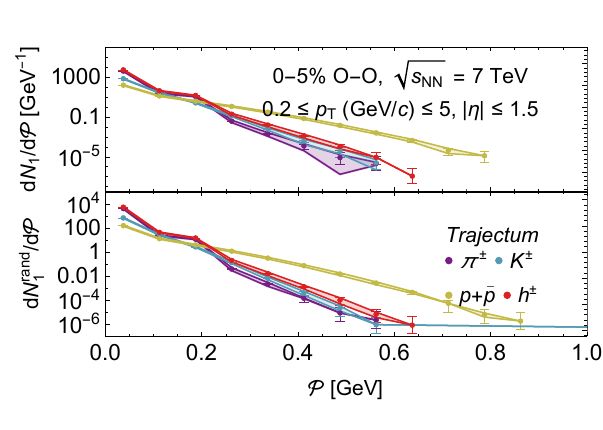}       \caption{Dimension-1 persistence distributions for pions, kaons, (anti)protons and all charged hadrons in the 0--5\% centrality class.
        Further details are those of \Cref{fig:persistencebetti0_05}. 
        The left panel shows data for collisions of \PbPb nuclei, the right panel for \OO nuclei.
        The top row depicts the plain persistence distributions $\dd N_1/\dd \Pcal$ and the bottom row has been computed for events with randomized azimuthal angles, $\dd N_1^{\mathrm{rand}}/\dd\Pcal$\@.}
    \label{fig:persistencebetti1_05_nobottom}
\end{figure*}
In \Cref{fig:persistencebetti0_05} the persistence distributions for homology dimension 0, together with the persistence distributions for randomized angles and the ratio between $\dd N^{\Delta}_0/\dd\Pcal$ and $\dd N_0/\dd\Pcal$, are shown for different particle species. 
The left and right panels display results in the 0--5\% centrality class for \PbPb and \OO collisions, respectively. 
For both \PbPb and \OO collisions, we observe a clear mass ordering in $\dd N_0/\dd\Pcal$ and $\dd N^{\textrm{rand}}_0/\dd\Pcal$ at very small persistences $\Pcal\simeq 0.05\textrm{~\GeVc}$, since in this regime the observable is multiplicity-dominated.
Above these small persistences, the persistence distributions for the different particle species closely follow each other in the case of \PbPb collisions, and only at large $\Pcal\gtrsim 0.8\textrm{~\GeVc}$ the curves begin to deviate slightly from each other.
For \OO collisions, the persistence distributions already differentiate among the species at intermediate persistences.
This is due to the different particle multiplicities among \PbPb and \OO and, therefore, differences among the inter-point distances in the constructed point clouds in both cases.

Concerning the impact of azimuthal correlations on the persistence distributions as detected by $(\dd N_0^\Delta/\dd \Pcal)/(\dd N_0/\dd\Pcal)$ (bottom row in \Cref{fig:persistencebetti0_05}), their influence is rather small, in contrast to the previously analyzed Betti curves (see \Cref{fig:betti0_Pb,fig:betti0_O,fig:betti1_05})\@.
In particular, the ratio $(\dd N_0^\Delta/\dd \Pcal)/(\dd N_0/\dd\Pcal)$ is dominated by statistical fluctuations for large persistence values.

In \Cref{fig:persistencebetti1_05_nobottom} the persistence distributions for homology dimension 1 along with the persistence distributions for randomized azimuthal angles are shown, analogously to before for the 0--5\% centrality class for \PbPb and \OO collisions. 
Again, we observe a mass ordering for small persistences. Subsequently, we observe saddle points in the distributions $\dd N_0/\dd\Pcal$ and $\dd N^{\textrm{rand}}_0/\dd\Pcal$ around $\Pcal\simeq 0.2\textrm{~\GeVc}$ for both \PbPb and \OO collisions.
These stem from the \pt selection, which we implemented in our analysis to be consistent with the momentum region accessible at the experiment. 
We note that for persistences $\Pcal\gtrsim 0.4\textrm{~\GeVc}$, dimension-1 holes appear only rarely, reflected by the persistence distributions.
For this reason, the ratio $(\dd N_0^\Delta/\dd \Pcal)/(\dd N_0/\dd\Pcal)$ is barely accessible given the statistics of our simulations.

\begin{figure*}[ht!]
    \centering
    \includegraphics[width=\linewidth]{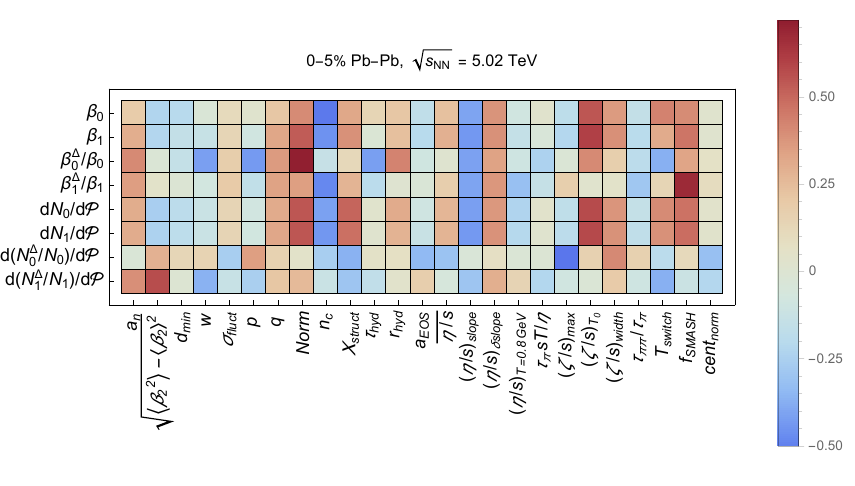}
    \includegraphics[width=\linewidth]{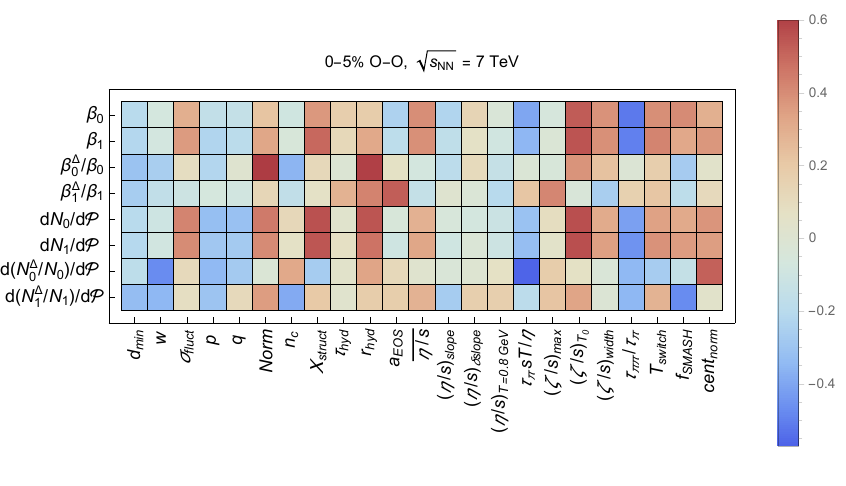}
    \caption{Pearson correlation coefficients of simulation parameters with different persistent homology observables for central \PbPb collisions (upper panel) and central \OO collisions (lower panel). 
    For $\beta_0(r)$, $\beta_1(r)$ and the relative difference ratios, the correlator are computed for homology radii $r=0.47\pm 0.04$ \GeVc as described in the main text\@. 
The persistence distributions are instead considered at $\mathcal P=0.10\pm0.04$~\GeVc\@.}
    \label{fig:pearson_central}
\end{figure*}
\clearpage
%%%%%%%%%%%%%%%%%%%%%%%%%%%%%%%%%%%%%%%%%%%%%%%%%%%%%%%%%%%%%%%%%%%%%%%%

%%%%%%%%%%%%%%%%%%%%%%%%%%%%%%%%%%%%%%%%%%%%%%%%%%%%%%%%%%%%%%%%%%%%%%%%
\subsection{Correlations}
In this section, we calculate the Pearson correlation coefficient, defined as
\begin{equation}
\rho = \frac{\sum_{i=1}^{n} (x_i - \bar{x})(y_i - \bar{y})}{\sqrt{\sum_{i=1}^{n} (x_i - \bar{x})^2 \sum_{i=1}^{n} (y_i - \bar{y})^2}},
\end{equation}
to quantify a possible (linear) relation between the tunable parameters of the simulation, $\{x_i\}$, and the observables $\{y_i\}$, here provided by the Betti curves. 
The tunable parameters, $\{x_i\}$, are derived from posterior distributions obtained through the Bayesian inference performed in \cite{Giacalone:2023cet}, which in principle --- computational constraints aside --- enables us to incorporate prior knowledge on the parameters' distribution 
and experimental constraints into the analysis. 
The computed correlation coefficients quantify the sensitivity of the persistent homology observables to changes in the underlying simulation parameters, allowing for an identification of the most influential parameters.

We do not subtract contributions from statistical uncertainties in this analysis.
This is because systematic uncertainties dominate, rendering the statistical contribution negligible in comparison.
The Pearson correlation coefficient is computed for 25 parameters in the case of \PbPb and 23 in the case of \OO collisions, that were all varied in the Bayesian analysis in \cite{Giacalone:2023cet}, as explained in the previous section (for further details see \Cref{Appendix:parameters})\@.

The Betti curves are analyzed in regions of homology radii where the sensitivity to parameter variations is most pronounced, and statistical uncertainties are negligible compared to systematic uncertainties. 
This is done by qualitatively looking at the spread of the 20 individual calculations across different observables and selecting the interval in which the spread is the widest. 
We note that the systematic uncertainties are always dominant, except for large homology radii.
For $\beta_0(r)$, $\beta_1(r)$ and the relative difference ratios, the correlator is computed for a homology radius $r=0.47\pm 0.04$ \GeVc\@. 
The persistence distributions are instead considered at $\mathcal P=0.10\pm0.04$ \GeVc\@. 

The results for the Pearson correlation coefficients in central \PbPb and \OO collisions are displayed in \Cref{fig:pearson_central} (see \Cref{Appendix:semicentral} for the results in semicentral collisions)\@. 
We observe that \PbPb and \OO show in most cases a similar pattern of (anti)correlations.
We observe clear indications of strong correlations between several tunable model parameters and the persistent homology observables. 
For instance, $\beta_0(r)$ shows a natural correlation with the \emph{Norm} parameter, as well as with the switching temperature $T_{\text{switch}}$, both of which influence the particle yields produced in the collisions.

\begin{figure}[ht]
    \centering
    \includegraphics[width=\linewidth]{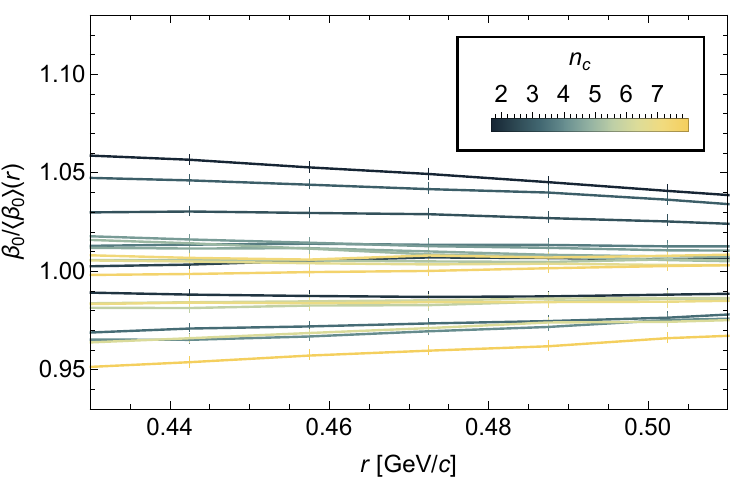}
    \caption{The dimension-0 Betti curve, $\beta_0(r)$, normalized by its average value is shown for different values of the $n_c$ parameter. }
    \label{fig:nc_1_1}
\end{figure}

However, some correlations may only be apparent. 
As an example, we investigate the unexpected correlation between the number of nucleon constituents $n_c$ and $\beta_0(r)$\@. 
In \Cref{fig:nc_1_1} we show $\beta_0(r)$ for different values of $n_c$ divided by the average $\langle\beta_0\rangle(r)$\@. 
We provide data for the range of radii $r=0.47 \pm 0.04$ \GeVc, for which the Pearson correlation coefficients have been computed. 
Although the Pearson coefficient indicates a strong anti-correlation, we cannot conclude that $\beta_0(r)$ and $n_c$ are linearly correlated. 
In fact, no color-ordering of $\beta_0$ is visible with decreasing $n_c$\@. 
The correlation is driven by outliers and therefore only apparent.

\begin{figure}[ht]
    \centering
    \includegraphics[width=\linewidth]{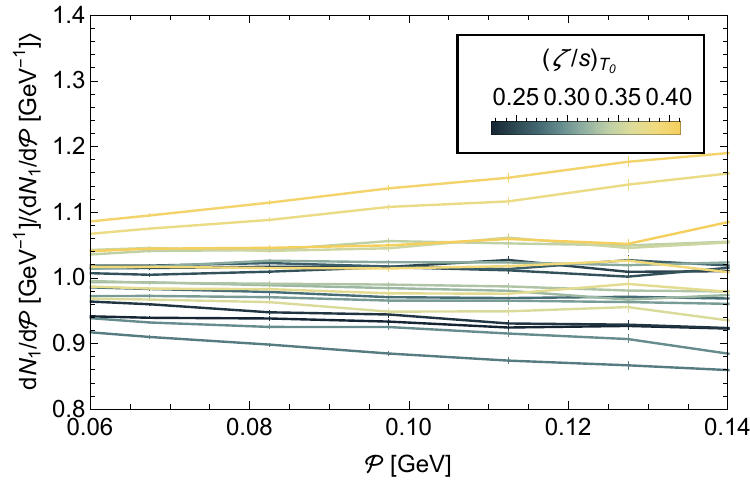}
    \caption{The curve $\dd N_1/\dd\Pcal$ normalized by its average value is shown for different values of the $(\zeta/s)_{T_0}$ parameter. }
    \label{fig:bulkT0_6_1}
\end{figure}

Among the genuine correlations, we highlight for example the one between the persistence distribution of $\dd N/\dd \mathcal{P}_1$ and the temperature at which the bulk viscosity has its maximum, $(\zeta/s)/T_0$, shown in \Cref{fig:bulkT0_6_1}\@. 
Such a correlation indicates that the measurement of $\dd N/\dd \mathcal{P}_1$, and the subsequent inclusion in the Bayesian analysis, might lead to a better constraint of the parameter $(\zeta/s)/T_0$\@.

\section{Conclusions}
\label{conclusion}

In this work, we studied simulated heavy-ion collisions via topological methods, in particular via Betti curves and persistence distributions.
We employed \trajectum, a state-of-the-art hydrodynamic model, to generate predictions that can serve as a benchmark for future experimental work.

First, we built a dictionary to explore the relationships between persistent homology-based and traditional observables such as particle multiplicities, momentum distributions and $n$-point connected correlation functions.

For instance, a connection between Betti curves of homology dimension 0 and particle multiplicities has been immediate. 
We highlighted the typical hydrodynamical feature of a mass-ordering for the Betti curves, which can be linked to the coupling of radial flow with the particle mass. 
An impact of anisotropic flow on the quotients $\beta_0^\Delta(r)/\beta_0(r)$ and $\beta_1^\Delta(r)/\beta_1(r)$ across different collision systems and centralities has been identified.
This showed that the randomization procedure has a larger impact where the anisotropic flow is supposed to be larger, namely in semicentral collisions. 
Whereas in this work we focused on the geometry of the transverse plane, a natural extension would be to study more complex persistent homology observables by incorporating the longitudinal direction.

However, we cannot conclude that persistent homology observables as the ones analyzed in this work effectively carry more information than is contained in the standard $n$-point correlators normally computed in heavy-ion physics analyses. 
At least in part, this can be inferred by looking at the Pearson correlation coefficients. 
Some of the Betti curves and persistence distributions display a strong correlation with the parameters employed for the Bayesian analysis with \trajectum. 
However, most of them have not been larger than the correlations observed for standard observables~\cite{Nijs:2021clz}.

When looking at the final results for the Betti curves, one can notice that the systematic error band rarely exceeds 20\% of the expectation values of the corresponding observables, and is compatible with the statistical uncertainty.
A larger systematic band could have indicated a potentially strong constraining power from such an observable, but this is not the case. 
Indeed, constraining power can be associated with an observable whose systematic uncertainty is larger compared to the uncertainties achieved experimentally.
Since the latter are at present not available for our observables, one could alternatively assess this by producing mock data, performing a Bayesian analysis, and checking whether the fit results improve with respect to the state of the art. 
We leave this analysis for future work. 

Ultimately, the constraining power of persistent homology observables can only be fully assessed through experimental measurements. 
If the theoretical models would exhibit significant deviations from the experimental measurements, Betti curves can prove highly valuable. 
In the absence of such measurements, further studies in this direction may yield limited insights. 

However, the introduction of new topological observables in the study of heavy-ion collisions introduces new avenues for their usage in novel analysis strategies that may prove valuable.
For instance, persistent homology typically probes long-range correlations in the system and can be well incorporated into supervised and unsupervised machine learning pipelines~\cite{hensel2021survey}.
In particular, the symbiosis of persistent homology with machine learning architectures can enhance the analysis sensitivity to hidden and extended structures in the underlying data.
Our approach to studying final states in simulations of heavy-ion collisions may therefore benefit from the application of topological machine learning.

%%%%%%%%%%%%%%%%%%%%%%%%%%%%%%%%%%%%%%%%%%%%%%%%%%%%%%%%%%%%%%%%%%%%%%%%

%%%%%%%%%%%%%%%%%%%%%%%%%%%%%%%%%%%%%%%%%%%%%%%%%%%%%%%%%%%%%%%%%%%
\begin{acknowledgments}
This work is funded by the Deutsche Forschungsgemeinschaft (DFG, German Research Foundation) under Germany’s Excellence Strategy EXC 2181/1 - 390900948 (the Heidelberg STRUCTURES Excellence Cluster) and the Collaborative Research Centre, Project-ID No. 273811115, SFB 1225 ISOQUANT.
\end{acknowledgments}

\appendix
\begin{figure*}[ht!]
    \centering
    \includegraphics[width=\linewidth]{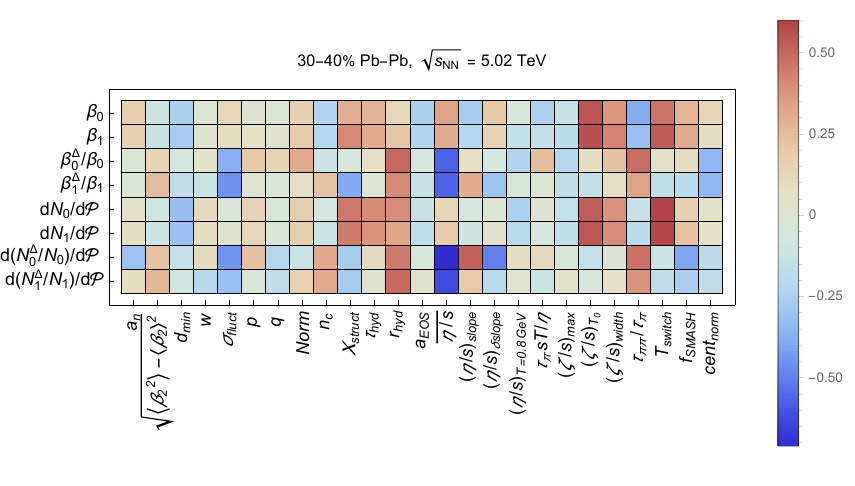}
    \includegraphics[width=\linewidth]{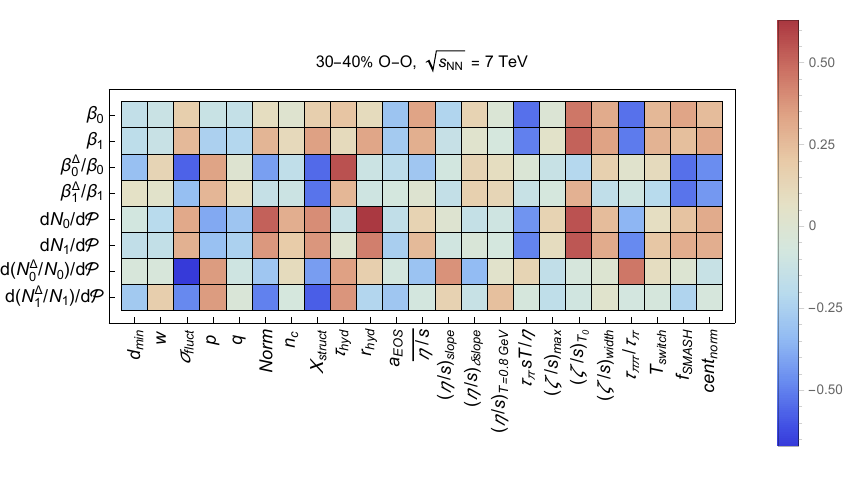}
    \caption{Pearson correlator of simulation parameters and persistent homology observables for semicentral \PbPb collisions (upper panel) and semicentral \OO collisions (lower panel).}
    \label{fig:pearson_semicentral}
\end{figure*}
\section{The mathematics of homology and persistent homology}\label{Appendix:Homology}
In this appendix we describe the construction of homology groups and persistent homology from a more mathematical perspective.
We first provide a brief introduction to simplicial and chain complexes, then give the mathematical construction of homology groups and finally of persistent homology.
While here we focus on a description of only the most relevant mathematical constructions, we refer to the literature for more details on homology and, more generally, algebraic topology, see e.g.~\cite{hatcher:2002, munkres2018elements}\@.
More comprehensive mathematical introductions to persistent homology are provided e.g.~by~\cite{otter2017roadmap, ChazalMichelIntro}\@.

\subsection{Simplicial and chain complexes}
An $\ell$-simplex $\sigma$ is a set of $\ell+1$ points, $\sigma = \{x_0,\ldots,x_\ell\}$\@.
We call a subset of $\sigma$ consisting of $\ell$ points a face of $\sigma$\@.
For instance, if $\sigma$ is a triangle, then all its edges are faces of $\sigma$\@.
A set $\mathcal{C}$ of simplices is called a simplicial complex, if (\emph{i}) for all simplices in $\mathcal{C}$ all their faces are also included in $\mathcal{C}$, and (\emph{ii}) for any pair of simplices $\sigma_1,\sigma_2$ with non-empty intersection, the latter is a face of both $\sigma_1$ and $\sigma_2$\@.
Examples of simplicial complexes are provided by the alpha complexes considered in the main text.

In order to define homology classes, one needs to have a generalized notion of paths of simplices.
This is accomplished by the construction of chain groups.
We consider chain complexes with coefficients in $\zz_2$, which will facilitate a particularly easy interpretation of their elements.
Then, an $\ell$-chain is a sequence of $\zz_2$-elements (i.e., $0$'s or $1$'s), one for each $\ell$-simplex in the simplicial complex $\mathcal{C}$.
For $\zz_2$-coefficients, we can therefore think of an $\ell$-chain as specifying for all $\ell$-simplices in $\mathcal{C}$ whether they are included in the chain or not.
The $\ell$-th chain complex $C_\ell(\mathcal{C})$ is the set of all such sequences.

Given the chain complexes for all relevant simplex dimensions, such as $C_0(\mathcal{C})$, $C_1(\mathcal{C})$, $C_2(\mathcal{C})$ for our 2-dimensional scenario, they can be related through the boundary operator.
Namely, we set 
\begin{equation}
    \partial_\ell: C_\ell(\mathcal{C})\to C_{\ell-1}(\mathcal{C})\,,
\end{equation}
so that $\partial_\ell$ maps an $\ell$-chain $c$ to the $(\ell-1)$-chain, which has a 1 whenever a specific $(\ell-1)$-simplex in $\mathcal{C}$ appears as a face of an uneven number of simplices in $c$, else 0\@.
For instance, if $c\in C_2(\mathcal{C})$ corresponds to two neighboring triangles with an edge in common, then $\partial_2 c$ corresponds to the collection of edges provided by the boundary of the union of the two triangles.
The face of the two triangles, which appears in the boundary of both of them, is not included in $\partial_2 c$.
Since boundaries of chain boundaries are empty, one has $\partial_{\ell-1}\circ \partial_\ell = 0$\@.
One thus obtains the chain complex
\begin{equation}
    \ldots \longrightarrow C_2(\mathcal{C})\overset{\partial_2}{\longrightarrow} C_1(\mathcal{C})\overset{\partial_1}{\longrightarrow} C_0(\mathcal{C}) \overset{0}{\longrightarrow} 0\,,
\end{equation}
where the composition of any two consecutive maps is zero.

\subsection{Homology groups}
In order to finally be able to define homology as provided by homology groups, we need to have a notion of closed paths of simplices.
For instance, a path along edges, i.e., a connected collection of edges, is characterized as closed, if it has an empty boundary.
If the path would not be closed, it would come with boundary points.
Formally, we say that a 1-chain $c\in C_1(\mathcal{C})$ is closed, if it is mapped to zero upon application of the first boundary operator: $\partial_1 c=0$.

These considerations can be generalized to an arbitrary simplex dimension.
We define the $\ell$-th cycle group $Z_\ell(\mathcal{C})$ as the kernel of the $\ell$-th boundary operator, i.e., the set of all those $\ell$-chains which get mapped to zero upon application of $\partial_\ell$:
\begin{equation}
Z_\ell(\mathcal{C}):=\ker (\partial_\ell)\,.    
\end{equation}
The elements of $Z_\ell$ are $\ell$-chains without boundaries and called cycles. 
They can indeed be thought of as closed paths of $\ell$-simplices in accordance with the earlier considerations for paths along edges.

For us, homology will be about holes and connected components in simplicial complexes.
For instance, in two dimensions, a hole can come about due to closed paths along edges (i.e., 1-cycles), which circumscribe triangles that are not part of the simplicial complex.
To describe what is \emph{not} a hole, we therefore need a notion of those cycles, which appear as boundaries of simplices in one simplex-dimension higher.
This can again be accomplished using the boundary operators.
We define the $\ell$-th boundary group $B_\ell(\mathcal{C})$ of $\mathcal{C}$ to consist of all those $\ell$-chains in $C_\ell(\mathcal{C})$, which are boundaries of $(\ell+1)$-chains:
\begin{equation}
B_\ell(\mathcal{C}):=\mathrm{im}(\partial_{\ell+1})\,.
\end{equation}

Crucially, if $c\in B_\ell(\mathcal{C})$, then we can write $c=\partial_{\ell+1} c'$, so $\partial_\ell c = \partial_\ell \partial_{\ell+1}c' = 0$.
Hence, $B_\ell(\mathcal{C})\subseteq Z_\ell(\mathcal{C})$ as subgroups, such that we can define their quotient groups,
\begin{equation}
H_\ell(\mathcal{C}):= Z_\ell(\mathcal{C})/B_\ell(\mathcal{C})\,.
\end{equation}
The $H_\ell(\mathcal{C})$ are finally called homology groups and their elements are $\ell$-dimensional homology classes. 
Homology classes correspond to sets of closed paths of $\ell$-simplices, defined modulo those $\ell$-cycles which appear as boundaries of $(\ell+1)$-simplices.
We may therefore think of homology classes as independent holes of dimension $\ell$.
Their number is the $\zz_2$-dimension of $H_\ell(\mathcal{C})$, called the $\ell$-th Betti number:
\begin{equation}
\beta_\ell(\mathcal{C}) := \dim_{\zz_2}(H_\ell(\mathcal{C}))\,,
\end{equation}
which we have already defined in the main text through a more intuitive approach.
This appendix provides the corresponding formal constructions.

One can show that homology classes remain invariant under continuous deformations of the simplicial complex $\mathcal{C}$\@.
This implies, for instance, local stretchings and compressions on the level of the complex, but in general not moving around the underlying point cloud elements.
We can therefore study the topology of $\mathcal{C}$ using the homology groups $H_\ell(\mathcal{C})$.
When $\mathcal{C}$ is considered as a topological space, its homology groups capture similar topological information compared to its homotopy groups, but the two are in general not the same.
They are related via Hurewicz's theorem, see e.g.~\cite{hatcher:2002}.

\subsection{Persistent homology}
At least in brevity, we now turn to persistent homology groups, whose mathematical constructions are a bit more abstract and involved than those leading to the homology groups.
Let $\{\mathcal{C}_r\}_{r\geq 0}$ be a filtration of simplicial complexes, i.e., a nested sequence of simplicial complexes so that for all $r\leq s$ we have $\mathcal{C}_r\subseteq \mathcal{C}_s$, for instance the alpha complex filtration considered in the main text.
One can compute all their individual homology groups $\{H_\ell(\mathcal{C}_r)\}_r$. 
In addition, the filtration provides us with inclusion maps $\mathcal{C}_r\hookrightarrow \mathcal{C}_s$ for all $r\leq s$. 
These induce certain maps on the homology groups:
\begin{equation}
\iota_\ell^{r,s}: H_k(\mathcal{C}_r)\to H_k(\mathcal{C}_s)\,.
\end{equation}
The map $\iota_\ell^{r,s}$ maps a homology class in $H_\ell(\mathcal{C}_r)$ either to the corresponding non-trivial homology class in $H_\ell(\mathcal{C}_s)$, if it is still present for $\mathcal{C}_s$, or to zero, if corresponding (potentially deformed) cycles appear as boundaries in $H_\ell(\mathcal{C}_s)$. 
Furthermore, non-trivial cokernels can appear for $\iota_\ell^{r,s}$: new homology classes may appear in $\mathcal{C}_s$, which are not present in $\mathcal{C}_r$. 
Then $s$ can be chosen such that for sufficiently small $\epsilon> 0$:
\begin{equation}\label{Eq:HomologyClassBirth}
H_\ell(\mathcal{C}_{s-\epsilon}) \subsetneq H_\ell(\mathcal{C}_s)\,.
\end{equation}
The collection of all homology groups put together with all induced maps between them, $\{(H_\ell(\mathcal{C}_r),\iota_\ell^{r,s})\}_{r\leq s}$, is called a persistence module. 
It is tame, if \Cref{Eq:HomologyClassBirth} holds for only finitely many distinct $s$-values. 

By the structure theorem of persistent homology~\cite{Edelsbrunner2000TopologicalPersistence,ZomorodianCarlsson2004ComputingPH}, any tame persistence module is isomorphic to its persistence diagram, i.e., the collection of all its birth-death pairs $(r_b,r_d)$, $r_b<r_d\in \rr\cup\{\infty\}$. 
The same birth-death pair may appear multiple times.
This provides the mathematical ground for the intuitively accessible interpretation of persistent homology employed in the main text.

\section{Correlations in semicentral collisions}

\label{Appendix:semicentral}
In \Cref{fig:pearson_semicentral} we show the results for the Pearson correlation coefficient in semicentral \PbPb and \OO collisions. We generally observe a similar pattern of positive and negative correlations in agreement with what is observed in central collisions (see \Cref{fig:pearson_central})\@. 
One difference worth highlighting is the strong correlation of the relative differences and relative difference persistence distribution of homology dimension 0 and 1 with the shear viscosity to entropy ratio ${\eta/s}$, which was completely absent in central collisions. 
This can be attributed to the high elliptic flow measured in non-central \PbPb collisions, which is expected from hydrodynamic calculations to be modulated by the shear viscosity. 
As we pointed out in the main text, the effect of the angular randomization, quantified by the quotients discussed here, is stronger where the azimuthal anisotropy of the point cloud is larger.

We observe that, similarly to central collisions, the Pearson correlation coefficients lie between (-0.6, 0.6). 
Although some of the Betti curves and persistence distributions display a strong correlation with the parameters employed for the Bayesian analysis with \trajectum, most of them are not larger than the correlations that can be observed for standard observables~\cite{Nijs:2021clz}\@. 
We notice that the surprising correlation observed in central \PbPb collisions between the Betti curves and the number of quark constituents of the nucleon $n_c$ is not found anywhere else, supporting our claim that it was only an apparent correlation.
However, we remark instead once again the strong correlation of several observables with the temperature at which the bulk viscosity to entropy ratio peaks, $(\zeta/s)_{T_0}$, which persists across all the analyzed systems and centralities, indicating that the inclusion of the Betti curves in a Bayesian analysis could potentially help in constraining better this fluid parameter.

\section{Parameters of \emph{Trajectum}}
\label{Appendix:parameters}

This appendix discusses the parameters used throughout the text.
The first two parameters concern the generation of Woods-Saxon nuclei such as ${}^{208}$Pb\@.
The $a_n$ parameter determines the thickness of the neutron skin, while $\sqrt{\langle\beta_2^2\rangle - \langle\beta_2\rangle^2}$ allows the quadrupole moment to vary on a nucleus-by-nucleus basis \cite{Giacalone:2023cet}\@.
The next parameter, $d_\text{min}$, determines the minimum internucleon distance, and is enforced both for ${}^{208}$Pb and ${}^{16}$O\@.
Each nucleon is then given $n_c$ constituents, with the constituent size determined by $\chi_\text{struct}$, and the total nucleon size given by $w$ \cite{Moreland:2018gsh}\@.

After generating the nuclei and the constituents themselves, an initial state is created through the T\raisebox{-0.5ex}{R}ENTo model.
T\raisebox{-0.5ex}{R}ENTo first generates two thickness functions (one for each nucleus) from the nucleon constituents, where the overall normalization is given by the Norm parameter, and the matter deposition from each nucleon can fluctuate, where $\sigma_\text{fluct}$ determines the size of these fluctuations \cite{Bernhard:2016tnd}\@.
The thickness functions are then combined by a formula with parameters $p$ and $q$ \cite{Nijs:2023yab}\@.

The initial state is then passed to the pre-hydrodynamic stage, which evolves the system up to proper time $\tau_\text{hyd}$\@.
The pre-hydrodynamic stage interpolates between free streaming and an AdS/CFT inspired model using the $r_\text{hyd}$ parameter, so that free streaming corresponds to $r_\text{hyd} = 0$ and AdS/CFT corresponds to $r_\text{hyd} = 1$\@.

At proper time $\tau_\text{hyd}$, the simulation is passed on to DNMR hydrodynamics.
We vary one equation of state parameter, $a_\text{EOS}$, and we have temperature dependent shear and bulk viscosities, parameterized by $\overline{\eta/s}$, $(\eta/s)_\text{slope}$, $(\eta/s)_{\delta\text{slope}}$ and $(\eta/s)_{T = 0.8\,\text{GeV}}$, and $(\zeta/s)_\text{max}$, $(\zeta/s)_\text{width}$ and $(\zeta/s)_{T_0}$, respectively.
For the precise functional form of the temperature dependence, see \cite{Nijs:2023yab}\@.
We also vary two second order transport coefficients, $\tau_\pi sT/\eta$, and $\tau_{\pi\pi}/\tau_\pi$\@.

Finally, at a temperature $T_\text{switch}$ the fluid is turned back into particles, and the simulation is continued with the SMASH code \cite{Weil:2016zrk,dmytro_oliinychenko_2020_3742965,Sjostrand:2007gs}\@.
In SMASH, all interactions are scaled by a factor $f_\text{SMASH}$, which we use to quantify the effect of potential deviations for some of the cross-sections that are used that are less precisely known.

The last parameter concerns the anchor point of where exactly 100\% centrality lies.
The $cent_\text{norm}$ parameter multiplies the anchor point, so that uncertainty on it is properly taken into account.

One can also notice that the number of parameters quoted is not always the same throughout the text.
This has two reasons.
Firstly, the $a_n$ and $\sqrt{\langle\beta_2^2\rangle - \langle\beta_2\rangle^2}$ parameters are used to create a Woods-Saxon nucleus, and as such they are not used in \OO simulations, which use explicit nuclear configurations instead of a Woods-Saxon parameterization.
This explains the 23 vs.~25 parameters that are varied in the present analysis.
The Bayesian analysis quoted, however, is stated as using 26 parameters.
The reason for this is that the Bayesian analysis fits to two collision energies, which doubles the `Norm' parameter, as it is different for each collision energy.

\bibliography{literature}

\end{document}

%% file: commands.tex
\newcommand{\pt}           {\ensuremath{p_{\rm T}}\xspace}
\newcommand{\px}           {\ensuremath{p_{\rm x}}\xspace}
\newcommand{\py}           {\ensuremath{p_{\rm y}}\xspace}

\newcommand{\sqrtsNN}      {\ensuremath{\sqrt{s_{\mathrm{NN}}}}\xspace}

\newcommand{\sevennn}       {$\sqrt{s_{\mathrm{NN}}}=7$~TeV\xspace}

\newcommand{\GeVc}         {GeV$/c$\xspace}